\documentclass[%
 amsmath,amssymb,
 aps,
 prd,
twocolumn, superscriptaddress
]{revtex4-2}

\usepackage[normalem]{ulem}

\usepackage{graphicx}
\usepackage{dcolumn}
\usepackage{bm}
\usepackage{hyperref}
\usepackage{amssymb}
\usepackage{xcolor}

\usepackage[dvipsnames]{xcolor}
\hypersetup{colorlinks,citecolor=OrangeRed}
\hypersetup{colorlinks=true, linkcolor=blue, urlcolor=MidnightBlue}

\begin{document}

\preprint{APS/123-QED}

\title{Magnetic pressure anisotropy in ultra-dense white dwarfs: Landau quantization and neutronization effects}

\author{Edson Otoniel}
 \email{edson.otoniel@ufca.edu.br}
 \affiliation{Instituto de Formaç\~ao de Educadores, Universidade Federal do Cariri, R. Oleg\'ario Emidio de Araujo, s/n – Aldeota, 63260-000 Brejo Santo, CE, Brazil
}

\author{Juan M. Z. Pretel}
 \email{juan04manuel91@gmail.com}
 \affiliation{Centro Brasileiro de Pesquisas F{\'i}sicas, Rua Dr.~Xavier Sigaud, 150 URCA, Rio de Janeiro CEP 22290-180, RJ, Brazil
}

\author{Victor B. T. Alves}
 \email{victor.bruno@discente.ufma.br}
 \affiliation{Programa de Pós-graduação em Física, CCET - Universidade Federal do Maranh\~ao, Campus Universit\'ario do Bacanga; CEP 65080-805, S\~ao Lu\'is, MA, Brasil}

\author{César O. V. Flores}
 \email{cesarovfsky@gmail.com}
 \affiliation{Centro de Ci\^encias Exatas, Naturais e Tecnol\'ogicas, CCENT - Universidade Estadual da Regi\~ao Tocantina do Maranh\~ao; C.P. 1300,\\ CEP 65901-480, Imperatriz, MA, Brasil
}
 \affiliation{Programa de Pós-graduação em Física, CCET - Universidade Federal do Maranh\~ao, Campus Universit\'ario do Bacanga; CEP 65080-805, S\~ao Lu\'is, MA, Brasil}

\author{Clésio E. Mota}
 \email{clesio200915@hotmail.com}
 \affiliation{Programa de Pós-graduação em Física, CCET - Universidade Federal do Maranh\~ao, Campus Universit\'ario do Bacanga; CEP 65080-805, S\~ao Lu\'is, MA, Brasil.
}

\date{\today}

\begin{abstract}
Strong magnetic fields modify the thermodynamics of white dwarf (WD) matter by quantizing the transverse motion of degenerate electrons into Landau levels and splitting the pressure into components parallel and perpendicular to the magnetic field. We investigate how this microscopic pressure anisotropy affects ultra-dense WD sequences when Coulomb lattice contributions, electron capture thresholds, neutron drip limits, and nuclear composition are treated consistently in the equation of state (EoS). The stellar equilibrium structure is obtained by solving an anisotropic extension of the Tolman--Oppenheimer--Volkoff (TOV) equations, where $P_r = P_\parallel$ and $P_t = P_\perp$ denote the radial and tangential pressures, respectively. We compare the post-electron-capture branches with their corresponding pre-capture sequences for pure $^{12}\rm{C}$, $^{16}\rm{O}$, $^{20}\rm{Ne}$, and $^{24}\rm{Mg}$ compositions, as well as for fixed 50/50 C/O, Ne/O, and Mg/Ne mixtures. The EoS predicts $P_\perp < P_\parallel$ in density ranges where only a few Landau levels are occupied. Consequently, pressure anisotropy shifts the stellar sequences toward smaller radii, with only minor changes in the maximum masses. We also find that post-capture configurations cannot be interpreted from the mass--radius plane alone, since the corresponding $M(\rho_c)$ curves reveal portions with $dM/d\rho_c<0$ that are flagged by the turning-point diagnostic after the maximum mass is reached. Our results show that strong magnetic fields significantly reduce the radii of low-mass WDs, placing the calculated branches in the same broad mass--radius domain as a selected observational sample, without constituting a specific observational identification or statistical validation, and suggesting that future high-precision observations of massive magnetic WDs could help assess the role of magnetic pressure anisotropy in determining their masses and radii.
\end{abstract}

\maketitle

\section{Introduction}\label{sec:intro}

Observational evidence indicates that magnetic fields constitute a common and relevant ingredient in the structure of WDs. In fact, spectroscopic surveys show that magnetic WDs exhibit a wide range of field strengths and are found both as isolated objects and in binary systems, with fields ranging from $10^{3}\,\rm{G}$ to $10^{9}\,\rm{G}$ \cite{Kepler2013, Ferrario2015, Landstreet2019, Bagnulo2022}. In interacting binaries, magnetic WDs can give rise to pulsed emission in the radio, optical, X-ray, and polarimetric bands, establishing white-dwarf pulsars and their binaries as sources of well-established observational results \cite{Marsh2016, Buckley2017, Pelisoli2023, Schwope2023, Rose2026}. Cyclotron radiation features and the presence of magnetic outflows further imply that strong fields can directly influence radiation mechanisms, angular-momentum loss, and the post-merger evolution of these objects \cite{Schmidt2001, vanRoestel2025, Kashiyama2019}. These results provide strong motivation for theoretical investigations of magnetized WDs, particularly within stellar models where magnetic stresses modify the hydrostatic equilibrium and generate anisotropic pressure. Magnetic fields may affect the observable properties of WDs not only through modifications of the microscopic EoS, but also through an effective anisotropic description, in which the effects of the magnetic field are encoded in the energy-momentum tensor rather than introduced via an explicit coupling between the magnetic field and gravity.

From a microscopic point of view, in a magnetized degenerate Fermi gas, the magnetic field selects a preferred spatial direction, so that the pressure parallel to the field differs from the transverse pressure, with the general relation $P_{\perp}=P_{\parallel}-\mathcal{M}B$, where $\mathcal{M}$ is the magnetization \cite{Strickland2012}. This mechanism has been explicitly demonstrated in magnetized quark matter, where Landau quantization and magnetization generate deviations between $P_{\parallel}$ and $P_{\perp}$ that become relevant for sufficiently strong fields \cite{Menezes2015}. In the context of magnetized WDs, such pressure splitting modifies the mass-radius relation and may suppress the emergence of stable super-Chandrasekhar configurations when the Maxwell and vacuum contributions are consistently included \cite{AlvearTerrero2015}. More generally, anisotropic pressures require stellar structure equations beyond the standard isotropic TOV framework, since the magnetic field induces an axially symmetric deformation of the star and may lead to a maximum field above which equilibrium configurations cease to exist \cite{Paret2015}. Relativistic Einstein-Maxwell calculations further show that magnetic fields can modify global observables such as radius, deformation, central density, and particle composition, even when the direct magnetization correction to the EoS has only a modest impact on the maximum mass \cite{Franzon2016}. Therefore, magnetic-field-induced anisotropy provides a physical channel through which the internal magnetic structure of WDs can affect macroscopic observables such as the mass-radius relation, stability limits, and deformation of highly magnetized compact stars.

The discovery of strongly magnetized WDs has opened new avenues for understanding the behavior of matter under extreme physical conditions. Magnetic fields ranging from $10^{10}$ to $10^{15}\,\rm{G}$ can drastically modify the phase space of the degenerate electron gas that provides the hydrostatic support for these compact objects. When the magnetic field strength becomes comparable to the quantum critical field (i.e., $B_c = 4.414 \times 10^{13}\,\rm{G}$), the classical continuum of electron momentum perpendicular to the magnetic field is quantized into discrete Landau levels. Such quantization restricts the available microstates and gives rise to an anisotropic pressure tensor. Consequently, the structural modeling of these stars requires a generalized EoS that distinctly treats the longitudinal pressure ($P_{\parallel}$) and the transverse pressure ($P_{\perp}$). 

Furthermore, at the extreme densities found in the cores of massive WDs, the electron Fermi energy becomes high enough to induce inverse beta decay (electron capture), transforming standard stable isotopes into exotic, neutron-rich nuclei. In that regard, we present a comprehensive EoS for magnetized WDs that consistently incorporates Landau quantization, Coulomb lattice interactions, and neutronization, and apply it to investigate the role of magnetic pressure anisotropy in stellar equilibrium.

Most investigations of anisotropic compact stars, including the seminal works of Bowers and Liang \cite{BowersLiang1974}, Horvat and collaborators \cite{Horvat2011}, Herrera and Barreto \cite{HerreraBarreto2013}, as well as many subsequent studies \cite{Raposo2019, Lau2024, Beltracchi2025, Guedes2026}, introduce the anisotropic pressure through phenomenological prescriptions chosen primarily for mathematical convenience, regularity requirements, or stability considerations \cite{Mak2003, Abreu2007, HerreraSantos1997}. In such models, the tangential pressure is typically specified through a phenomenological ansatz rather than derived from the underlying microscopic physics of dense matter. In light of this, a substantial body of work has shown that anisotropic stresses can have profound implications for the structure and dynamical stability of quark stars and neutron stars. Interestingly, even when introduced through purely phenomenological profiles, the anisotropic pressure has been found to significantly affect the mass-radius relation and tidal deformability \cite{Rahmansyah2020, Das2023, Pretel:2023nlr, Khunt2026}, rotational properties~\cite{Pattersons:2021lci, Beltracchi2024, Becerra2024, Beltracchi2025}, and the spectrum of both radial \cite{Pretel2020EPJC, Mohanty2024, Becerra2026} and nonradial oscillations~\cite{Doneva2012, Curi2022, Arbanil:2023yil, Mohanty:2023hha, Pretel2024JCAP, Arbanil2026}. Although observational evidence for anisotropic pressure is still inconclusive, it has recently been recognized as a valuable diagnostic of missing physics in compact-star models \cite{Pang2026}, motivating its investigation in strongly magnetized WDs.

Recent studies of magnetized neutron stars have shown that pressure anisotropy can arise naturally from the underlying microphysics through the interaction of matter with strong magnetic fields \cite{Bordbar2022}. In that regard, the anisotropy considered in the present work is not introduced phenomenologically, but emerges from a well-defined physical mechanism. In strongly magnetized degenerate matter, the Landau quantization of electron motion breaks the isotropy of momentum space, leading to distinct pressure components parallel and perpendicular to the magnetic field \cite{Canuto1968, Perez2008}. As a consequence, the anisotropic stress tensor follows directly from the microscopic energy spectrum of the electron gas and can be computed self-consistently from the EoS itself, rather than being prescribed through an arbitrary functional form. This provides a physically identifiable origin for the anisotropy in WDs and removes the need to postulate an \emph{a priori} anisotropy profile $\sigma(r)$.

The paper is organized as follows. In Sec.~\ref{sec:methods} we construct the magnetized WD EoS, including the electron gas, Coulomb lattice, composition dependence, and weak interaction thresholds. In Sec.~\ref{sec:TOVequations} we introduce the effective anisotropic stellar structure equations used to obtain the equilibrium sequences. In Sec.~\ref{sec:results} we present the post-electron-capture branch, pure composition, mixed composition, and observational context results. Finally, Sec.~\ref{sec:discussion} summarizes the main physical implications and outlines the limitations and future extensions of the model. Throughout this manuscript, we use physical units, except in Sec.~\ref{sec:TOVequations}, where the stellar structure equations are formulated in geometric units.

\section{Equation of state of magnetized white-dwarf matter}
\label{sec:methods}

The matter sector considered in this work is a cold, fully ionized WD plasma composed of nuclei and degenerate electrons in a locally uniform magnetic field. The magnetic field defines a preferred direction at the microscopic level. As a consequence, the electron contribution to the stress tensor is not described by a single scalar pressure, but by a pressure parallel to the field, \(P_\parallel\), and a pressure perpendicular to it, \(P_\perp\) \cite{Canuto1968,Strickland2012,Perez2008}. The local matter equation constructed below provides the thermodynamic quantities that are later supplied to the stellar-structure equations. It does not by itself determine the global magnetic-field geometry.

The calculation is performed at zero temperature. The ions are treated as a static Coulomb lattice and their thermal pressure is neglected. The ionic rest-mass contribution is retained in the energy density, while the electron gas supplies the dominant degeneracy pressure. We write the composition in multicomponent form from the outset. A nuclear species \(j\) is specified by charge number \(Z_j\), mass number \(A_j\), ion number density \(n_j\), and experimentally determined atomic mass \(M_{\rm atom}(A_j,Z_j)\). Local charge neutrality gives
\begin{equation}\label{eq:wd_charge_neutrality}
    n_e = \sum_j Z_j n_j ,
\end{equation}
where $n_e$ is the electron number density. The baryon number density and electron fraction are
\begin{equation}
    n_B = \sum_j A_j n_j,
    \qquad
    Y_e=\frac{n_e}{n_B}.
    \label{eq:wd_nb_ye}
\end{equation}
The mass density associated with the ionic rest masses is
\begin{equation}
    \rho = \sum_j n_j M_{\rm atom}(A_j,Z_j).
    \label{eq:wd_rho_atomic_mass}
\end{equation}
For a one-component plasma, Eqs.~(\ref{eq:wd_charge_neutrality}) to (\ref{eq:wd_rho_atomic_mass}) reduce to
\begin{align}
    n_e &= Zn_i, \\
    n_B &= An_i, \\
    Y_e &= \frac{Z}{A}, \\
    \rho &= n_iM_{\rm atom}(A,Z)
    = n_e\frac{M_{\rm atom}(A,Z)}{Z}.
    \label{eq:wd_one_component_limit}
\end{align}
The atomic masses are taken from experimental nuclear data whenever available and converted consistently to CGS units. These masses play the same physical role as in standard crust calculations based on atomic mass evaluations \cite{Wang2021AME,Chamel2014}.

The rest-mass part of the material energy density must be combined with the explicitly calculated free-electron energy density without double counting the electron rest masses already included in the neutral atomic mass. For a one-component layer, the implemented rest-energy contribution is therefore
\begin{equation}
    \epsilon_{\rm rest}
    =
    n_i M_{\rm atom}(A,Z)c^2
    -
    n_e m_ec^2 ,
    \label{eq:wd_rest_energy}
\end{equation}
where \(m_e\) is the electron mass. In multicomponent notation this becomes
\begin{equation}
    \epsilon_{\rm rest}
    =
    \sum_j n_j M_{\rm atom}(A_j,Z_j)c^2
    -
    n_e m_ec^2 .
    \label{eq:wd_multicomponent_rest_intro}
\end{equation}
This term is supplemented by the magnetized electron energy density and by the Coulomb lattice energy density, as described below.

\subsection{Magnetized electron gas}
\label{subsec:wd_magnetized_electrons}

The electron critical magnetic field is
\begin{equation}
    B_c=\frac{m_e^2c^3}{e\hbar}
    \simeq 4.414\times10^{13}\,{\rm G},
    \label{eq:wd_bc}
\end{equation}
and the dimensionless field strength is defined by
\begin{equation}\label{eq:wd_bstar}
    B_\star=\frac{B}{B_c}.
\end{equation}
The reduced Compton wavelength of the electron is
\begin{equation}
    \lambda_e=\frac{\hbar}{m_ec}.
    \label{eq:wd_lambda_e}
\end{equation}
For a uniform magnetic field oriented along the \(z\)-axis, the transverse electron motion is quantized into Landau levels, while the momentum \(p_z\) parallel to the field remains continuous. Neglecting the anomalous magnetic moment, the electron energy spectrum is
\begin{equation}
    E_\nu(p_z) =
    \left[
    c^2p_z^2
    +
    m_e^2c^4\left(1+2\nu B_\star\right)
    \right]^{1/2},
    \quad
    \nu=0,1,2,\ldots
    \label{eq:wd_landau_energy}
\end{equation}
The spin degeneracy factor is
\begin{equation}
    g_\nu=
    \begin{cases}
    1, & \nu=0,\\
    2, & \nu\ge 1 .
    \end{cases}
    \label{eq:wd_landau_degeneracy}
\end{equation}
Writing the electron chemical potential as
\begin{equation}
    \gamma_e=\frac{\mu_e}{m_ec^2},
    \label{eq:wd_gamma_e}
\end{equation}
the dimensionless longitudinal Fermi momentum in level \(\nu\) can be written as
\begin{equation}
    x_\nu =
    \left(\gamma_e^2-1-2\nu B_\star\right)^{1/2}.
    \label{eq:wd_xnu}
\end{equation}

Only levels with real $x_\nu$ are occupied. The maximum occupied Landau level is consequently
\begin{equation}
    \nu_{\max} =
    \left\lfloor\frac{\gamma_e^2-1}{2B_\star}\right\rfloor .
    \label{eq:wd_numax}
\end{equation}

The $B=0$ reference curves are evaluated with the isotropic zero-field limit of the EoS; the value $B_\star=0$ is not inserted directly into the Landau-level sums. The electron number density is obtained by summing the longitudinal Fermi lengths of all occupied levels, namely
\begin{equation}
    n_e(\gamma_e,B) =
    \frac{B_\star}{2\pi^2\lambda_e^3}
    \sum_{\nu=0}^{\nu_{\max}} g_\nu x_\nu .
    \label{eq:wd_ne_landau}
\end{equation}
For a specified matter density and composition, Eq.~(\ref{eq:wd_rho_atomic_mass}) fixes the ion densities and Eq.~(\ref{eq:wd_charge_neutrality}) fixes \(n_e\). Equation~(\ref{eq:wd_ne_landau}) is then inverted for \(\gamma_e\) at fixed \(B\).

It is useful to introduce
\begin{equation}
    a_\nu = 1+2\nu B_\star .
    \label{eq:wd_anu}
\end{equation}
The electron energy density is
\begin{equation}
    \epsilon_e
    =
    \frac{m_ec^2}{\lambda_e^3}
    \frac{B_\star}{4\pi^2}
    \sum_{\nu=0}^{\nu_{\max}}
    g_\nu
    \left[
    \gamma_e x_\nu
    +
    a_\nu
    \ln\left(
    \frac{\gamma_e+x_\nu}{\sqrt{a_\nu}}
    \right)
    \right].
    \label{eq:wd_electron_energy}
\end{equation}
The pressure parallel to the field is
\begin{equation}
    P_{\parallel,e}
    =
    \frac{m_ec^2}{\lambda_e^3}
    \frac{B_\star}{4\pi^2}
    \sum_{\nu=0}^{\nu_{\max}}
    g_\nu
    \left[
    \gamma_e x_\nu
    -
    a_\nu
    \ln\left(
    \frac{\gamma_e+x_\nu}{\sqrt{a_\nu}}
    \right)
    \right].
    \label{eq:wd_p_parallel_e}
\end{equation}
This expression is the longitudinal pressure associated with the matter grand potential,
\begin{equation}
    \Omega_e = \epsilon_e-\mu_en_e = -P_{\parallel,e}.
    \label{eq:wd_grand_potential}
\end{equation}

On the other hand, the perpendicular electron pressure is evaluated from the transverse matter stress,
\begin{equation}
    P_{\perp,e}
    =
    \frac{m_ec^2}{\lambda_e^3}
    \frac{B_\star^2}{2\pi^2}
    \sum_{\nu=0}^{\nu_{\max}}
    g_\nu \nu
    \ln\left(
    \frac{\gamma_e+x_\nu}{\sqrt{a_\nu}}
    \right).
    \label{eq:wd_p_perp_e}
\end{equation}
Equivalently, the pressure splitting may be written in terms of the electron magnetization,
\begin{equation}
    \mathcal{M}_e =
    -\left(\frac{\partial \Omega_e}{\partial B}\right)_{\mu_e}
    =
    \left(\frac{\partial P_{\parallel,e}}{\partial B}\right)_{\mu_e},
    \label{eq:wd_magnetization}
\end{equation}
so that
\begin{equation}
    P_{\perp,e}
    =
    P_{\parallel,e}
    -
    \mathcal{M}_e B .
    \label{eq:wd_pperp_magnetization}
\end{equation}
Equations~(\ref{eq:wd_p_parallel_e}) and (\ref{eq:wd_p_perp_e}) are evaluated as explicit sums over the occupied Landau levels. In the lowest-Landau-level limit, \(\nu_{\max}=0\), the sum in Eq.~(\ref{eq:wd_p_perp_e}) vanishes because the summand is proportional to \(\nu\). Thus, the transverse electron pressure is suppressed in that limiting matter contribution, whereas the parallel pressure remains finite. This statement refers to the electron matter pressure and does not by itself imply a global collapse of the star.

\begin{figure*}[t]
    \centering
    \includegraphics[width=\textwidth]{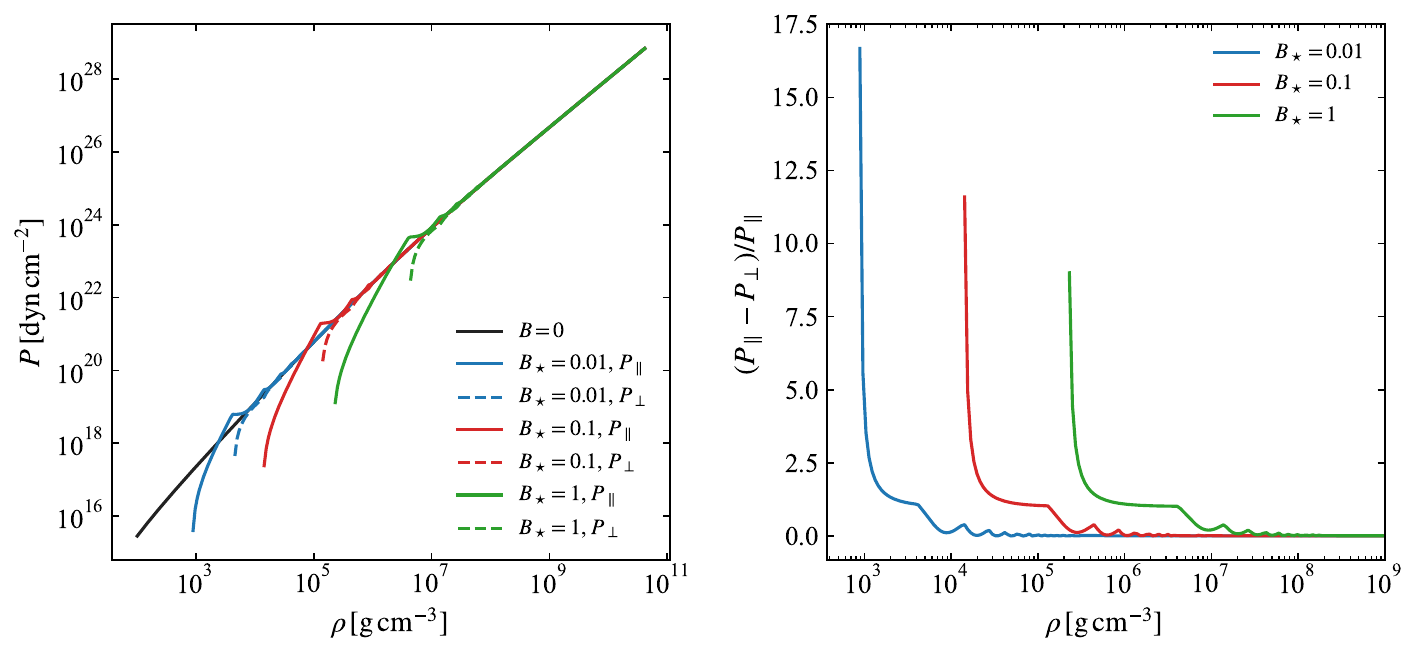}
    \caption{Material pressure branches and pressure anisotropy for pure \(^{12}\mathrm{C}\) matter. The left panel compares $P_\parallel$ (by solid curves) and $P_\perp$ (by dashed lines) as functions of mass density for three values of $B_\star$ \eqref{eq:wd_bstar}, together with a \(B=0\) isotropic reference. The right panel displays the relative anisotropy \((P_\parallel-P_\perp)/P_\parallel\) for the magnetized cases. Notice that only positive pressure values are displayed on the logarithmic pressure axis.}
    \label{fig:eos_pressure_anisotropy}
\end{figure*}

In Fig.~\ref{fig:eos_pressure_anisotropy}, the main result is that the magnetic field affects the pressure balance most strongly in the low-density region, where only a small number of Landau levels is populated. In such a regime, the transverse material pressure is strongly suppressed relative to \(P_\parallel\), and the ratio \((P_\parallel-P_\perp)/P_\parallel\) can become of order unity or larger when the lattice correction drives \(P_\perp\) below the parallel branch. As the density increases, additional Landau levels are populated and the two pressure branches rapidly approach one another, so that the matter becomes effectively isotropic at high density. The oscillatory structure in the anisotropy curve is the de Haas--van Alphen response associated with successive Landau-level filling. Increasing \(B_\star\) shifts the anisotropic regime to higher densities, showing that stronger fields preserve a sizeable pressure splitting deeper inside the star. The $B=0$ curve is included as a benchmark to facilitate comparison between the isotropic pressure and the magnetized branches.

\subsection{Coulomb lattice and total material equation}
\label{subsec:wd_lattice_total_eos}

The ions form a Coulomb lattice embedded in the degenerate electron background. For a multicomponent lattice, the Coulomb energy density may be written in the form
\begin{equation}
    \epsilon_L
    =
    C_{\rm L}e^2 n_e^{4/3}
    \mathcal{F}\left(\{Z_j\},\{x_j\}\right),
    \label{eq:wd_lattice_energy_multi}
\end{equation}
where \(C_{\rm L}\) is the lattice constant for the chosen crystal structure, \(x_j=n_j/\sum_k n_k\) denotes the ionic fraction, and \(\mathcal{F}\) contains the charge and composition dependence of the Coulomb lattice. This notation makes explicit that the lattice correction depends on all ionic species present in a mixed layer, not only on a single charge number. For a one-component body-centered cubic lattice, Eq.~(\ref{eq:wd_lattice_energy_multi}) reduces to
\begin{equation}
    \epsilon_L
    =
    C_{\rm bcc}e^2 n_e^{4/3} Z^{2/3},
    \label{eq:wd_lattice_energy_ne}
\end{equation}
or, using charge neutrality, one can write it as
\begin{equation}
    \epsilon_L
    =
    C_{\rm bcc}e^2 Z^2 n_i^{4/3}.
    \label{eq:wd_lattice_energy_ni}
\end{equation}
Here \(C_{\rm bcc}=-1.444231\) is the body-centered cubic lattice constant in the convention adopted for Coulomb crystals \cite{Chamel2014}. For a binary mixture with species \((Z,n)\) and \((Z',n')\), charge neutrality reads
\begin{equation}
    n_e = Zn + Z'n',
    \label{eq:wd_binary_charge_neutrality}
\end{equation}
which is the two-component specialization of Eq.~(\ref{eq:wd_charge_neutrality}). The corresponding lattice factor can be written as a function of both charges and both abundances,
\begin{equation}
    \begin{aligned}
    \mathcal{F}_{2}
    &=
    \bar Z^{-4/3}
    \left[
    \eta Z^2
    +
    \zeta {Z'}^2
    +
    (1-\eta-\zeta)ZZ'
    \right],
    \\
    \bar Z
    &=
    \xi Z + (1-\xi)Z' .
    \end{aligned}
    \label{eq:wd_binary_lattice_factor}
\end{equation}
where \(\xi\) is the fraction of the first ionic species and \(\eta\), \(\zeta\) are constants fixed by the adopted binary lattice geometry. The corresponding lattice pressure is given by
\begin{equation}
    P_L=\frac{1}{3}\epsilon_L .
    \label{eq:wd_lattice_pressure}
\end{equation}
Since \(C_{\rm bcc}<0\), the lattice lowers the material energy density and gives a negative pressure correction.

The total material energy density is
\begin{equation}
    \epsilon_{\rm total}
    =
    \epsilon_{\rm rest}
    +
    \epsilon_e
    +
    \epsilon_L .
    \label{eq:wd_epsilon_total}
\end{equation}
The lattice correction is treated as isotropic in the matter-pressure contribution and, consequently, the total pressures are
\begin{equation}
    P_\parallel
    =
    P_{\parallel,e}
    +
    P_L ,
    \label{eq:wd_p_parallel_total}
\end{equation}
and
\begin{equation}
    P_\perp
    =
    P_{\perp,e}
    +
    P_L
    =
    P_{\parallel,e}
    -
    \mathcal{M}_eB
    +
    P_L .
    \label{eq:wd_p_perp_total}
\end{equation}
No separate pure Maxwell term \(B^2/(8\pi)\) is added to Eq.~(\ref{eq:wd_epsilon_total}) or to the pressure branches considered in this work. The quantities in Eqs.~(\ref{eq:wd_epsilon_total}) to (\ref{eq:wd_p_perp_total}) are therefore material thermodynamic quantities. If a global stellar model includes the electromagnetic stress explicitly, that contribution must be added consistently in the structure equations.

This omission is deliberate: the present calculation isolates the pressure anisotropy generated by the magnetized matter EoS and does not attempt to construct a global magnetic-field geometry. In a local field-aligned decomposition, the direct electromagnetic sector would contribute \(\epsilon_{\rm EM}=B^2/(8\pi)\), \(P_{\parallel}^{\rm EM}=-B^2/(8\pi)\), and \(P_{\perp}^{\rm EM}=+B^2/(8\pi)\). These terms would modify both the energy density and the two pressure components, while the associated Lorentz forces would require an axisymmetric Einstein--Maxwell treatment. Their inclusion would therefore change the mass--radius relations and generally introduce deformation, but the quantitative shift cannot be inferred from the present spherical matter-only calculation.

\subsection{Electron capture, composition sequences, and neutron drip}
\label{subsec:wd_capture_drip}

At sufficiently high density, the electron chemical potential can make a nucleus unstable against electron capture. In this work, electron capture is not introduced through a reaction rate. It is treated as a local thermodynamic threshold based on a Gibbs free-energy comparison at fixed pressure, following the standard crust-equilibrium logic used in dense compact-star matter \cite{Baym1971BPS,Chamel2014}. For a parent layer \((A,Z)\), the effective capture chain keeps the mass number fixed and lowers the charge according to the restricted reaction network. The chains considered here include
\begin{align}
    ^{12}{\rm C}:&\quad Z=6\rightarrow4, \nonumber\\
    ^{16}{\rm O}:&\quad Z=8\rightarrow6, \nonumber\\
    ^{20}{\rm Ne}:&\quad Z=10\rightarrow8\rightarrow6, \nonumber\\
    ^{24}{\rm Mg}:&\quad Z=12\rightarrow10\rightarrow8 .
    \label{eq:wd_capture_chains}
\end{align}
These chains define the finite composition space explored in the present calculation. They should not be interpreted as a complete cold-catalyzed minimization over all nuclei.

For each trial pre-capture state, the parent Gibbs quantity per baryon is computed as
\begin{equation}
    g^- =
    \frac{\epsilon^-+P_\parallel^-}{n_B^-}.
    \label{eq:wd_gibbs_parent}
\end{equation}
The daughter state is constructed with the next nucleus in the prescribed chain. Its electron density is obtained by imposing pressure continuity across the transition,
\begin{equation}
    P_\parallel^+(n_e^+)
    =
    P_\parallel^-(n_e^-).
    \label{eq:wd_pressure_matching_beta}
\end{equation}
The post-capture Gibbs quantity is then
\begin{equation}
    g^+ =
    \frac{\epsilon^+ + P_\parallel^+}{n_B^+},
    \label{eq:wd_gibbs_daughter}
\end{equation}
and the local stability margin is
\begin{equation}
    \Delta g_\beta = g^+ - g^- .
    \label{eq:wd_delta_g_beta}
\end{equation}
The parent composition is stable against that capture channel when \(\Delta g_\beta>0\). The transition threshold is defined by
\begin{equation}
    \Delta g_\beta = 0 .
    \label{eq:wd_beta_threshold}
\end{equation}
This pressure-matched construction keeps the lattice correction, atomic rest-energy term, and magnetized electron contribution on both sides of the transition. The pressure used for the thermodynamic matching is \(P_\parallel\), because it is tied to the matter grand-potential relation in Eq.~(\ref{eq:wd_grand_potential}). This convention is local to the EoS construction and does not by itself prescribe the radial pressure used in a stellar-structure calculation.

At these electron-capture transitions, the pressure-matching condition keeps $P_\parallel$ continuous while the parent and daughter densities may differ. The resulting density jumps are part of the piecewise EoS used in the capture-following sequences and do not represent a separate self-bound interface.

The no-free-neutron EoS is terminated when neutron drip is reached. The restricted channel used to diagnose the onset is
\begin{equation}
    (A,Z)+e^-
    \rightarrow
    (A-1,Z-1)+n+\nu_e .
    \label{eq:wd_drip_channel}
\end{equation}
The post-drip state is obtained by pressure matching in the same way as for electron capture, but the daughter state includes one free neutron per daughter ion at onset. At the threshold, the neutron contribution to the pressure is neglected, while the neutron rest-mass energy is included. If \(n_n\) is the free-neutron number density at onset, the post-drip Gibbs quantity is written as
\begin{equation}
    g_{\rm drip}^{+}
    =
    \frac{
    \epsilon_{\rm ion}^{+}
    +
    \epsilon_e^{+}
    +
    \epsilon_L^{+}
    +
    n_n m_nc^2
    +
    P_\parallel^{+}
    }
    {n_{B,{\rm ion}}^{+}+n_n},
    \label{eq:wd_gibbs_drip_plus}
\end{equation}
where \(m_n\) is the neutron mass. The neutron-drip margin is
\begin{equation}
    \Delta g_{\rm drip}
    =
    g_{\rm drip}^{+}
    -
    g^- ,
    \label{eq:wd_delta_g_drip}
\end{equation}
and the threshold is determined by
\begin{equation}
    \Delta g_{\rm drip}=0 .
    \label{eq:wd_drip_threshold}
\end{equation}
States beyond this threshold would require free-neutron degrees of freedom and are outside the validity domain of the present no-free-neutron matter equation.

The sequence is assembled by comparing, within each active layer, the next electron-capture threshold with the neutron-drip threshold. The event reached first sets the upper boundary of the layer. If electron capture occurs first, the matter equation continues with the post-capture nucleus at the pressure-matched density. If neutron drip occurs first, the no-free-neutron sequence stops. The local thermodynamic state is specified by
\begin{equation}
    \left\{
    \rho,\,
    n_e,\,
    n_B,\,
    B,\,
    B_\star,\,
    \epsilon_{\rm total},\,
    P_\parallel,\,
    P_\perp,\,
    A,\,
    Z,\,
    Y_e
    \right\},
    \label{eq:wd_local_variables}
\end{equation}
with the active nuclear layer determined by the capture and drip conditions. This structure makes it possible to compare pure pre-capture compositions, threshold-following equilibrium sequences, and controlled parent-daughter mixtures. The latter are used only as sensitivity models for partially converted layers unless a separate equilibrium calculation is supplied.

\begin{figure*}[t]
    \centering
    \includegraphics[width=\textwidth]{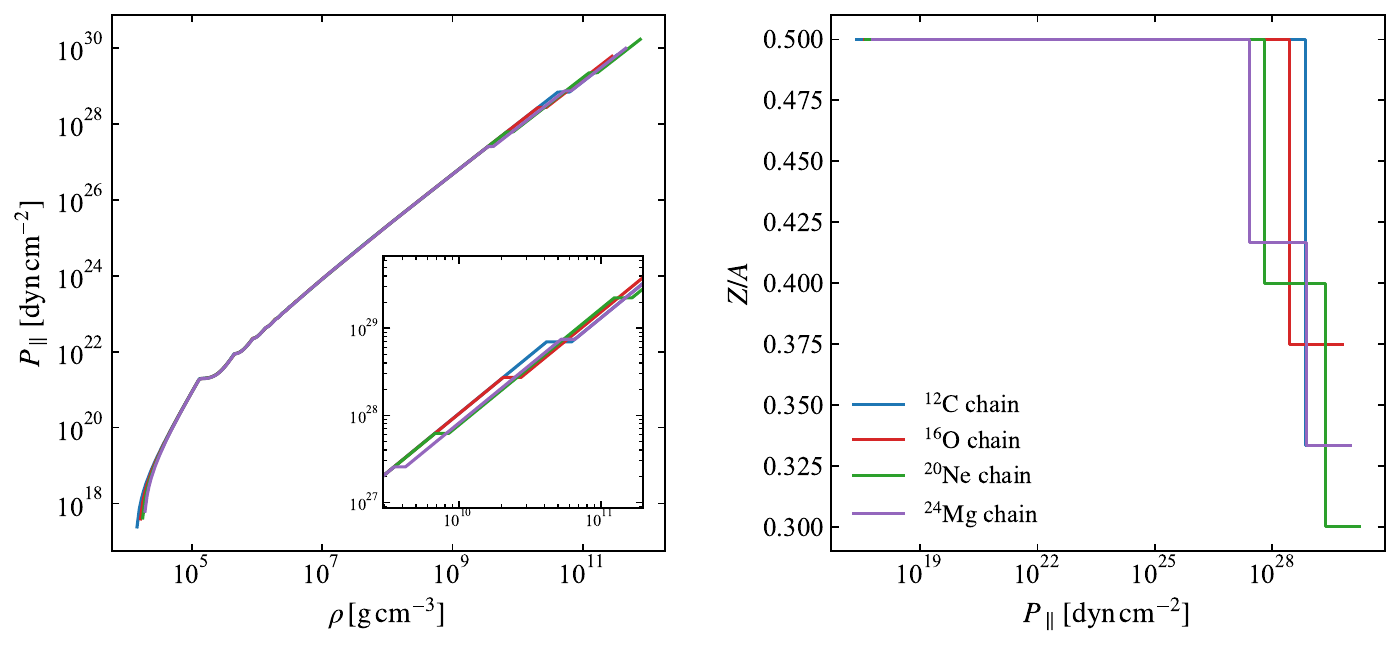}
    \caption{Pressure and composition changes along the electron-capture sequences at dimensionless field $B_\star=0.1$. The left panel shows \(P_\parallel(\rho)\), with an inset zooming the transition region. The right panel shows the stepwise reduction of \(Z/A\) as a function of \(P_\parallel\), making explicit the composition changes induced by electron capture before the sequence terminates at neutron drip.}
    \label{fig:eos_capture_sequence}
\end{figure*}

Table~\ref{tab:wd_thresholds_magnetic} and Fig.~\ref{fig:eos_capture_sequence} show that the first electron-capture instability is reached at lower density for larger initial nuclear charge in the sequence considered here. Thus \(^{24}\mathrm{Mg}\) captures before \(^{20}\mathrm{Ne}\), \(^{16}\mathrm{O}\), and \(^{12}\mathrm{C}\), reflecting the smaller stability window of the heavier parent nuclei against increasing electron chemical potential. Over the range \(B_\star=0.01\) to \(1\), the threshold densities change only weakly, indicating that these capture and drip thresholds are controlled mainly by the nuclear mass differences and the electron chemical potential rather than by the magnetic correction alone. The stepwise drops in \(Z/A\) in Fig.~\ref{fig:eos_capture_sequence} make explicit that each capture reduces the electron fraction at nearly continuous pressure, producing density jumps between parent and daughter layers. The neutron-drip entries mark the end of the no-free-neutron description; beyond them, the matter composition would require free-neutron degrees of freedom and cannot be represented by the present WD lattice sequence. For the \(50/50\) mixtures, the tabulated \(\rho_{\rm th}\) is the first inverse-beta threshold reached by either component of the mixture, not a neutron-drip density. These mixed thresholds therefore define the upper density of the controlled pre-capture mixed branches used later in the stellar-structure calculation.

\begin{table*}[t]
\caption{Electron-capture and neutron-drip thresholds for the composition sequences used in this work. The threshold density \(\rho_{\rm th}\) is the parent-side density immediately before the transition and is given in \(\mathrm{g\,cm^{-3}}\). For binary \(50/50\) mixtures, \(\rho_{\rm th}\) denotes the first inverse-beta threshold among the two components of the mixture. The last column gives the corresponding threshold pressure at \(B_\star=0.1\), in \(\mathrm{dyn\,cm^{-2}}\), as a representative pressure scale. The calculations include the Coulomb lattice contribution and use pressure matching with \(P_\parallel\).}
\label{tab:wd_thresholds_magnetic}
\begin{ruledtabular}
\begin{tabular}{llcccc}
Sequence & Event & \(\rho_{\rm th}(0.01)\) & \(\rho_{\rm th}(0.1)\) & \(\rho_{\rm th}(1)\) & \(P_{\rm th}(0.1)\) \\
\(^{12}\mathrm{C}\) & \(^{12}\mathrm{C}\rightarrow{}^{12}\mathrm{Be}\) & \(4.158{\times}10^{10}\) & \(4.158{\times}10^{10}\) & \(4.158{\times}10^{10}\) & \(6.992{\times}10^{28}\) \\
\(^{12}\mathrm{C}\) & neutron drip from \(^{12}\mathrm{Be}\) & \(3.275{\times}10^{11}\) & \(3.275{\times}10^{11}\) & \(3.275{\times}10^{11}\) & \(6.392{\times}10^{29}\) \\
\(^{16}\mathrm{O}\) & \(^{16}\mathrm{O}\rightarrow{}^{16}\mathrm{C}\) & \(2.059{\times}10^{10}\) & \(2.059{\times}10^{10}\) & \(2.059{\times}10^{10}\) & \(2.729{\times}10^{28}\) \\
\(^{16}\mathrm{O}\) & neutron drip from \(^{16}\mathrm{C}\) & \(2.817{\times}10^{11}\) & \(2.817{\times}10^{11}\) & \(2.817{\times}10^{11}\) & \(6.107{\times}10^{29}\) \\
\(^{20}\mathrm{Ne}\) & \(^{20}\mathrm{Ne}\rightarrow{}^{20}\mathrm{O}\) & \(6.811{\times}10^{9}\) & \(6.811{\times}10^{9}\) & \(6.812{\times}10^{9}\) & \(6.213{\times}10^{27}\) \\
\(^{20}\mathrm{Ne}\) & \(^{20}\mathrm{O}\rightarrow{}^{20}\mathrm{C}\) & \(1.247{\times}10^{11}\) & \(1.247{\times}10^{11}\) & \(1.247{\times}10^{11}\) & \(2.240{\times}10^{29}\) \\
\(^{20}\mathrm{Ne}\) & neutron drip from \(^{20}\mathrm{C}\) & \(7.639{\times}10^{11}\) & \(7.639{\times}10^{11}\) & \(7.639{\times}10^{11}\) & \(1.713{\times}10^{30}\) \\
\(^{24}\mathrm{Mg}\) & \(^{24}\mathrm{Mg}\rightarrow{}^{24}\mathrm{Ne}\) & \(3.515{\times}10^{9}\) & \(3.515{\times}10^{9}\) & \(3.516{\times}10^{9}\) & \(2.560{\times}10^{27}\) \\
\(^{24}\mathrm{Mg}\) & \(^{24}\mathrm{Ne}\rightarrow{}^{24}\mathrm{O}\) & \(5.276{\times}10^{10}\) & \(5.276{\times}10^{10}\) & \(5.276{\times}10^{10}\) & \(7.489{\times}10^{28}\) \\
\(^{24}\mathrm{Mg}\) & neutron drip from \(^{24}\mathrm{O}\) & \(4.568{\times}10^{11}\) & \(4.568{\times}10^{11}\) & \(4.568{\times}10^{11}\) & \(9.913{\times}10^{29}\) \\
\colrule
\multicolumn{6}{c}{Binary \(50/50\) mixtures: first inverse-beta threshold}\\
\(^{12}\mathrm{C}/^{16}\mathrm{O}\) & first beta in \(^{16}\mathrm{O}\) & \(2.057{\times}10^{10}\) & \(2.057{\times}10^{10}\) & \(2.057{\times}10^{10}\) & \(2.729{\times}10^{28}\) \\
\(^{20}\mathrm{Ne}/^{16}\mathrm{O}\) & first beta in \(^{20}\mathrm{Ne}\) & \(6.805{\times}10^{9}\) & \(6.805{\times}10^{9}\) & \(6.805{\times}10^{9}\) & \(6.213{\times}10^{27}\) \\
\(^{24}\mathrm{Mg}/^{20}\mathrm{Ne}\) & first beta in \(^{24}\mathrm{Mg}\) & \(3.513{\times}10^{9}\) & \(3.512{\times}10^{9}\) & \(3.513{\times}10^{9}\) & \(2.560{\times}10^{27}\) \\
\end{tabular}
\end{ruledtabular}
\end{table*}

\section{Stellar structure equations}\label{sec:TOVequations}

The internal structure of strongly magnetized WDs may be substantially modified by the anisotropic stresses generated by intense magnetic fields. In particular, the magnetic field introduces a preferred spatial direction within the stellar medium, thereby breaking the local isotropy of the fluid pressure. As a consequence, the matter distribution cannot be adequately described within the framework of an isotropic perfect fluid, making it necessary to adopt an anisotropic fluid formalism. Under these considerations, the energy-momentum tensor describing the magnetized stellar matter is written as \cite{BowersLiang1974, Horvat2011, Pretel:2023nlr}
\begin{equation}
T_{\mu\nu}
= (\varepsilon +P_\perp)u_\mu u_\nu + P_\perp g_{\mu\nu} - (P_\perp-P_\parallel) \chi_\mu \chi_\nu ,
\end{equation}
where $\varepsilon$ is the total energy density entering the stress tensor, while $P_\parallel$ and $P_\perp$ denote the radial and transverse pressures, respectively. The mass-density variable $\rho$ used in the EoS and in the tabulated central densities is converted to the corresponding $\varepsilon$ before the structure equations are integrated. The four-velocity $u^\mu$ satisfies the normalization condition
\begin{equation}
    u_\mu u^\mu=-1,
\end{equation}
while $\chi^\mu$ is a unit spacelike vector orthogonal to the fluid flow, so that 
\begin{equation}
    \chi_\mu \chi^\mu=1,
\qquad u_\mu \chi^\mu=0.
\end{equation}

In the static limit, appropriate for WDs in hydrostatic equilibrium, the matter-energy variables and metric functions depend exclusively on the radial coordinate. Consequently, the energy-momentum tensor becomes diagonal, namely
\begin{equation}
    T^\mu_{\ \nu} = \mathrm{diag}(-\varepsilon, P_\parallel, P_\perp, P_\perp).
\end{equation}
This form explicitly reflects the existence of anisotropic stresses inside the stellar medium. In the notation used for the anisotropic TOV equations, the identification is $P_r\equiv P_\parallel$ and $P_t\equiv P_\perp$; the parallel/perpendicular notation is retained for the microscopic EoS.

Assuming spherical symmetry as a first approximation, the interior spacetime of the star is described by the static metric
\begin{equation}\
ds^2 = -e^{2\psi(r)}dt^2 + e^{2\lambda(r)}dr^2 + r^2 d\Omega^2,
\end{equation}
where $d\Omega^2 = d\theta^2+\sin^2\theta\, d\phi^2$ is the line element on the unit two-sphere. The metric functions $\psi(r)$ and $\lambda(r)$ encode the gravitational field generated by the stellar energy-momentum content. It is important to emphasize that we adopt an effective anisotropic-fluid description in a spherically symmetric spacetime. In this approach, the macroscopic effects of the magnetic field are incorporated through pressure anisotropy, rather than by solving the coupled Einstein-Maxwell equations.

The field strengths explored in the stellar sequences are $B_\star=0.01$, $0.1$, and $1$, corresponding to $B\simeq4.414\times10^{11}$, $4.414\times10^{12}$, and $4.414\times10^{13}\,\mathrm{G}$, respectively. This interval is the parameter domain sampled in the effective spherical model we are adopting, not a quantitatively established range of validity. A reliable upper limit for the spherical approximation would require comparison with self-consistent axisymmetric Einstein--Maxwell solutions.

The gravitational field is determined by Einstein's equations,
\begin{equation}\label{FieldEqs}
G_{\mu\nu} = R_{\mu\nu} -\frac{1}{2}g_{\mu\nu}R
= 8\pi T_{\mu\nu},
\end{equation}
where geometrized units (i.e., $G=c=1$) are employed in this section. Combining Eq.~\eqref{FieldEqs} with the conservation law $\nabla_\mu T^{\mu\nu}=0$, one obtains the generalized TOV equations governing the hydrostatic equilibrium of anisotropic compact stars:
\begin{align}
\frac{dm}{dr}
&= 4\pi r^2 \varepsilon,  \label{TOV1}  \\
\frac{dP_\parallel}{dr}
&= -(\varepsilon +P_\parallel)
\left[ \frac{m+4\pi r^3 P_\parallel}{r(r-2m)} \right]
+
\frac{2}{r}(P_\perp-P_\parallel),  \label{TOV2}  \\
\frac{d\psi}{dr}
&= -\frac{1}{\varepsilon +P_\parallel}\frac{dP_\parallel}{dr}
+ \frac{2(P_\perp-P_\parallel)}{r(\varepsilon +P_\parallel)}. \label{TOV3}
\end{align}
Equation~\eqref{TOV2} clearly shows the role of anisotropy in the stellar equilibrium, where the additional term $2(P_\perp-P_\parallel)/r$ acts as a force contribution arising from the difference between transverse and radial pressures. In particular, positive anisotropy ($P_\perp > P_\parallel$) generates an outward-directed contribution that can partially counterbalance gravity. Conversely, negative anisotropy enhances the effective inward force. The dynamical stability of the resulting configurations cannot be determined from this equilibrium equation alone.

The mass function $m(r)$ represents the gravitational mass enclosed within a sphere of radius $r$, while the metric potential $\lambda(r)$ is determined through
\begin{equation*}
e^{-2\lambda(r)} = 1-\frac{2m(r)}{r}.
\end{equation*}
The surface is determined by the condition $P_\parallel(R)=0$, which defines the radius $R$ of the stellar configuration, while the total gravitational mass is given by $M = m(R)$.

In order to close the above system of differential equations, it is necessary to specify the EoSs associated with the parallel and perpendicular pressure components, namely $P_\parallel = P_\parallel(\rho)$ and $P_\perp = P_\perp(\rho)$. In the present work, these relations are derived from the thermodynamic description of a magnetized degenerate electron gas, which provides an appropriate microscopic framework for WD matter subjected to intense magnetic fields. Once the central mass density $\rho_c$ is given as an input, the system of Eqs.~\eqref{TOV1}-\eqref{TOV3} can be integrated outward from the stellar center by imposing the regularity conditions
\begin{equation}\label{BoundConditions}
m(0)=0,  \qquad  \varepsilon(0)=\varepsilon_c.
\end{equation}
Here $\varepsilon_c$ is obtained from the chosen central mass density $\rho_c$ through the material EoS. Finally, continuity of the interior metric with the exterior Schwarzschild spacetime requires
\begin{equation}
\psi(R) = \frac{1}{2} \ln\left(1-\frac{2M}{R}\right),
\end{equation}
which completes the set of boundary conditions defining the relativistic equilibrium problem for magnetized anisotropic WDs in Einstein gravity.

\section{Results}\label{sec:results}

The numerical results are organized around three increasingly controlled uses of the magnetized EoS. We first examine sequences that follow electron captures, in which the composition is allowed to change at the capture thresholds. This case exposes the consequence of density jumps and changes in \(Y_e\), but it is not the cleanest way to define a single metastable family of WDs. We then restrict the calculation to pure compositions terminated at the first beta instability threshold. These branches represent configurations before capture at fixed nuclear species. Finally, we construct binary \(50/50\) mixtures, truncated at the first capture instability of either component, to test whether an intermediate composition can connect the pure composition limits to C/O, Ne/O, and Mg/Ne material. Throughout this section, the mass versus radius curves are computed with the effective anisotropic prescription \(P_r=P_\parallel\) and \(P_t=P_\perp\), so the magnetic pressure splitting directly enters the stellar structure equation through the anisotropic term. Given an EoS, the anisotropic TOV equations, Eqs.~\eqref{TOV1} and \eqref{TOV2}, are solved numerically over a range of central densities using the boundary conditions \eqref{BoundConditions}. Each choice of the central density $\rho_c$ yields a stellar configuration characterized by its radius $R$ and mass $M$, thereby constructing the corresponding mass--radius diagram for our magnetized WDs. The central macroscopic question is whether this anisotropic pressure balance produces a measurable compactification of the stellar sequence relative to the isotropic reference or to the weak magnetic field reference.

\subsection{Sequences following captures and stability}

\begin{figure*}
    \centering
    \includegraphics[width=\textwidth]{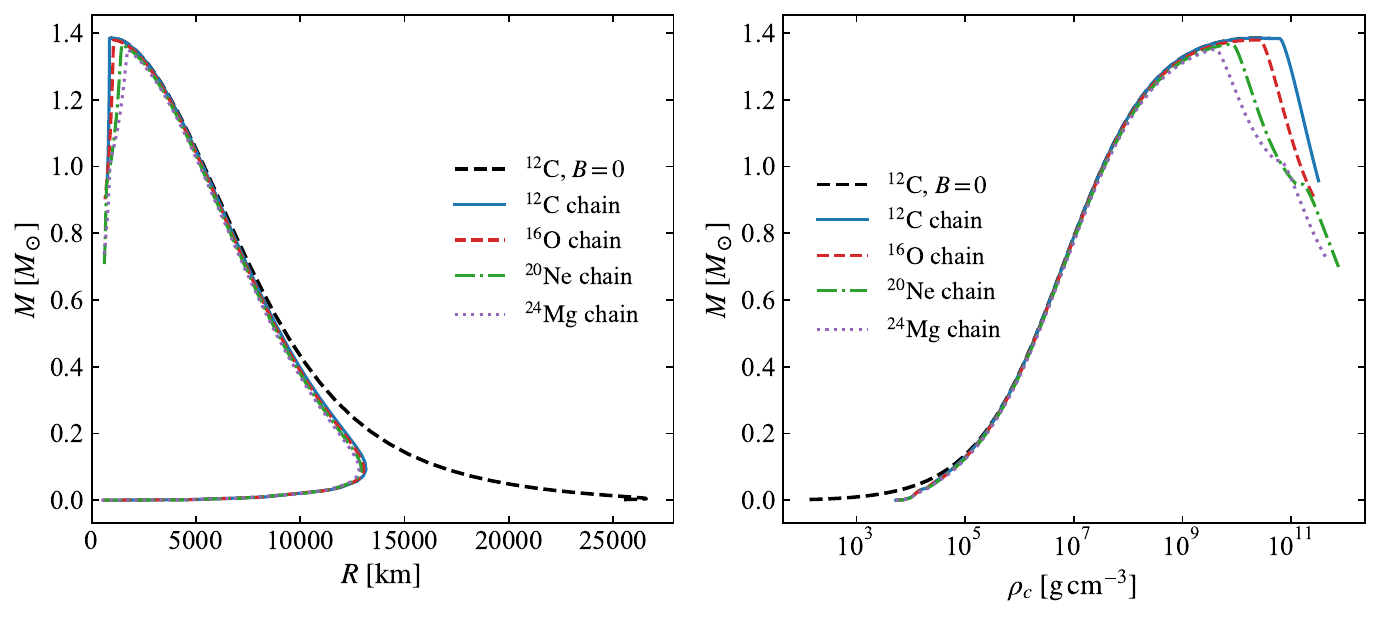}
    \caption{Mass versus radius relation and mass versus central density relation for sequences following captures at \(B_\star=0.01\), using the anisotropic TOV framework. The dashed black curve represents the $B=0$ carbon model, which serves as the reference benchmark for comparison. The magnetized curves follow the electron capture chains generated from the corresponding initial parent nuclei.}
    \label{fig:results_capture_sequence_tov}
\end{figure*}

Figure~\ref{fig:results_capture_sequence_tov} shows that the construction following captures does not behave as a single smooth stable family. The mass versus radius panel alone could suggest that the branches remain continuous enough to be used as ordinary stellar sequences. The \(M\) versus \(\rho_c\) panel is more diagnostic: after each curve reaches its maximum mass, the sampled sequence develops an extended region with \(dM/d\rho_c<0\). This is the relevant turning point criterion for radial stability in the effective spherical calculation. For the sequence initiated by carbon at \(B_\star=0.01\), the maximum mass is \(1.3849\,M_\odot\) at \(\rho_c=2.08\times10^{10}\,\mathrm{g\,cm^{-3}}\) and \(R=1051\,\mathrm{km}\). The sequence initiated by oxygen reaches \(1.3791\,M_\odot\) at \(\rho_c=2.02\times10^{10}\,\mathrm{g\,cm^{-3}}\), while the sequences initiated by neon and magnesium reach their maxima at lower central densities, \(8.58\times10^{9}\,\mathrm{g\,cm^{-3}}\) and \(4.25\times10^{9}\,\mathrm{g\,cm^{-3}}\), respectively. This ordering reflects the lower capture thresholds of the heavier parent nuclei already visible in the EoS analysis.

We use the sign change in $dM/d\rho_c$ only as a turning-point screening indicator for the effective spherical one-parameter sequences. No radial perturbation or eigenfrequency analysis is performed here, so the configurations are not labeled dynamically stable on this basis alone.

\subsection{Pure branches before capture and anisotropic compactification}

\begin{figure}
    \centering
    \includegraphics[width=\columnwidth]{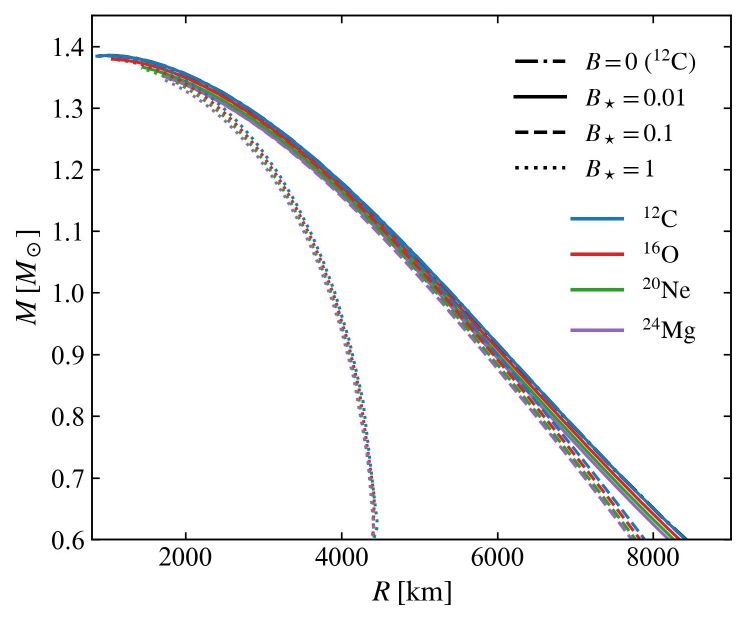}
    \caption{Mass versus radius curves for pure compositions before capture. Each branch is truncated at the first electron capture instability of the corresponding parent nucleus. The \(B=0\) curve is shown for the carbon reference, where the pressure is purely isotropic. The magnetized branches, on the other hand, are obtained from the anisotropic TOV equations, with $P_\parallel$ and $P_\perp$ identified as the radial and tangential pressures, respectively. }
    \label{fig:results_pure_mr}
\end{figure}

The preceding result motivates a more conservative interpretation. Instead of treating the sequence after capture as a single equilibrium family, one can define a metastable branch for a fixed nuclear composition and terminate it at the first electron capture threshold. This is the construction shown in Fig.~\ref{fig:results_pure_mr}. In this diagnostic version, the integration of the anisotropic TOV equations requires the radial pressure \(P_\parallel\) to be positive, while the tabulated transverse pressure is kept as the tangential stress. Here, ``metastable'' denotes a fixed-composition pre-capture construction and does not constitute a dynamical stability claim.

The most important macroscopic effect in Fig.~\ref{fig:results_pure_mr} is the compactification produced by the anisotropic prescription. At fixed composition, the stronger field curves are displaced toward smaller radii within the displayed mass range. This behavior follows directly from the sign of the anisotropic term in the effective TOV equations, $2(P_\perp-P_\parallel)/r$. Since the microscopic EoS gives \(P_\perp<P_\parallel\) in the density interval where the magnetic splitting is relevant, the anisotropic contribution is negative and steepens the radial pressure gradient. Within this radial effective model, the star therefore reaches a given mass at a smaller radius. This compactification is not a full axisymmetric Einstein--Maxwell result; it is the consequence of feeding the local microscopic pressure anisotropy into the spherical anisotropic structure equations.

The pure branches also show how the capture threshold limits the accessible endpoint at high density. The carbon branch extends to the largest central density, while the magnesium branch terminates much earlier. Consequently, the maximum mass decreases along the sequence from carbon to magnesium. At \(B_\star=0.01\), the maximum masses are \(1.3850\,M_\odot\), \(1.3791\,M_\odot\), \(1.3660\,M_\odot\), and \(1.3500\,M_\odot\) for \(^{12}\mathrm{C}\), \(^{16}\mathrm{O}\), \(^{20}\mathrm{Ne}\), and \(^{24}\mathrm{Mg}\), respectively. From \(B_\star=0.01\) to \(B_\star=1\), the maximum mass changes only weakly for carbon and oxygen, but the radii and endpoints are more visibly affected for the heavier branches. For example, the \(^{24}\mathrm{Mg}\) maximum changes from \(1.3500\,M_\odot\) at \(R=1744\,\mathrm{km}\) for \(B_\star=0.01\) to \(1.3467\,M_\odot\) at \(R=1704\,\mathrm{km}\) for \(B_\star=1\).

\subsection{Fixed binary mixtures}

The pure composition branches are useful limiting cases, but WDs are not generally expected to be chemically pure. To test an intermediate description without introducing a full reaction network, we generated binary mixtures with fixed ion fractions \(\xi=0.5\) in the implemented binary bcc lattice model. The mixtures considered were \(^{12}\mathrm{C}/^{16}\mathrm{O}\), \(^{20}\mathrm{Ne}/^{16}\mathrm{O}\), and \(^{24}\mathrm{Mg}/^{20}\mathrm{Ne}\). For each mixture, the equation was truncated at the first beta instability threshold among the two individual capture channels. This prescription does not model the subsequent chemical evolution of the mixture. It only defines a controlled mixed branch before capture.

The mixed sequences must therefore be interpreted as fixed-composition pre-capture equilibrium sequences, not as complete weak-equilibrium crust sequences. Once the first inverse-beta threshold is reached, the present calculation does not follow the subsequent captures, compositional redistribution, or reaction-network evolution.

\begin{figure}
    \centering
    \includegraphics[width=\columnwidth]{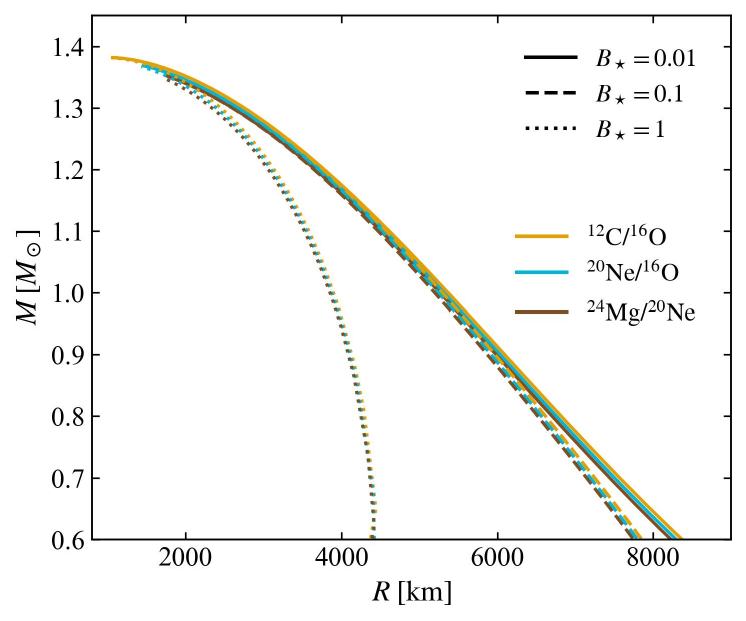}
    \caption{Mass versus radius curves for binary \(50/50\) mixtures truncated at the first beta instability channel. As the magnetic field strength increases, the radii of low-mass magnetized WDs systematically decrease, with a qualitative behavior similar to the chemically pure one shown in Fig.~\ref{fig:results_pure_mr}. }
    \label{fig:results_mixed_mr}
\end{figure}

The mixed curves in Fig.~\ref{fig:results_mixed_mr} fall between the corresponding pure composition limits. The C/O mixture reaches \(M_{\max}=1.3816\,M_\odot\) for \(B_\star=0.01\), lower than pure carbon but close to pure oxygen. The Ne/O mixture reaches \(1.3686\,M_\odot\), and the Mg/Ne mixture reaches \(1.3523\,M_\odot\). These values reflect the fact that the limiting density of the mixed branch is set by the first unstable component. In the C/O mixture the first channel is capture in the oxygen component, while in the Ne/O and Mg/Ne mixtures the first channel is capture in the heavier component. Thus the mixed branches are not arbitrary interpolations of the pure curves. Their endpoints are determined by the Gibbs threshold of the most vulnerable component in the mixture.

The same compactification trend seen in the pure branches is also present in the mixed cases. For each fixed mixture, the \(B_\star=1\) branch lies at smaller radii than the branches at weaker magnetic field in the plotted mass interval. The effect is clearest away from the maximum mass point, where the stronger field dotted curves move inward relative to the \(B_\star=0.01\) curves. The radius at maximum mass of the C/O mixture changes from \(1060\,\mathrm{km}\) at \(B_\star=0.01\) to \(1050\,\mathrm{km}\) at \(B_\star=1\), while the Ne/O mixture changes from \(1456\,\mathrm{km}\) to \(1431\,\mathrm{km}\). The Mg/Ne mixture changes from \(1746\,\mathrm{km}\) to \(1706\,\mathrm{km}\). These shifts quantify, in the present model, the contraction associated with the anisotropic pressure balance.

\begin{table*}
\caption{Maximum masses and corresponding radii for pure branches before capture and binary \(50/50\) mixed branches. All magnetized entries use \(P_r=P_\parallel\) and \(P_t=P_\perp\). The mixed branches are truncated at the first beta instability channel among the two components.}
\label{tab:results_maximum_masses}
\begin{ruledtabular}
\begin{tabular}{lcccc}
Composition & \(B_\star\) & \(M_{\max}/M_\odot\) & \(R(M_{\max})\,[\mathrm{km}]\) & \(\rho_c(M_{\max})\,[\mathrm{g\,cm^{-3}}]\)\\
\colrule
\multicolumn{5}{c}{Pure branches before capture}\\
\(^{12}\mathrm{C}\) & 0.01 & 1.3850 & 998 & \(2.48\times10^{10}\)\\
\(^{16}\mathrm{O}\) & 0.01 & 1.3791 & 1059 & \(2.02\times10^{10}\)\\
\(^{20}\mathrm{Ne}\) & 0.01 & 1.3660 & 1454 & \(6.68\times10^{9}\)\\
\(^{24}\mathrm{Mg}\) & 0.01 & 1.3500 & 1744 & \(3.45\times10^{9}\)\\
\(^{12}\mathrm{C}\) & 0.1 & 1.3850 & 1037 & \(2.18\times10^{10}\)\\
\(^{16}\mathrm{O}\) & 0.1 & 1.3791 & 1059 & \(2.02\times10^{10}\)\\
\(^{20}\mathrm{Ne}\) & 0.1 & 1.3660 & 1454 & \(6.68\times10^{9}\)\\
\(^{24}\mathrm{Mg}\) & 0.1 & 1.3500 & 1744 & \(3.45\times10^{9}\)\\
\(^{12}\mathrm{C}\) & 1 & 1.3846 & 993 & \(2.45\times10^{10}\)\\
\(^{16}\mathrm{O}\) & 1 & 1.3786 & 1050 & \(2.02\times10^{10}\)\\
\(^{20}\mathrm{Ne}\) & 1 & 1.3643 & 1429 & \(6.68\times10^{9}\)\\
\(^{24}\mathrm{Mg}\) & 1 & 1.3467 & 1704 & \(3.45\times10^{9}\)\\
\colrule
\multicolumn{5}{c}{Binary \(50/50\) mixed branches}\\
\(^{12}\mathrm{C}/^{16}\mathrm{O}\) & 0.01 & 1.3816 & 1060 & \(2.02\times10^{10}\)\\
\(^{20}\mathrm{Ne}/^{16}\mathrm{O}\) & 0.01 & 1.3686 & 1456 & \(6.67\times10^{9}\)\\
\(^{24}\mathrm{Mg}/^{20}\mathrm{Ne}\) & 0.01 & 1.3523 & 1746 & \(3.44\times10^{9}\)\\
\(^{12}\mathrm{C}/^{16}\mathrm{O}\) & 0.1 & 1.3816 & 1060 & \(2.02\times10^{10}\)\\
\(^{20}\mathrm{Ne}/^{16}\mathrm{O}\) & 0.1 & 1.3686 & 1455 & \(6.67\times10^{9}\)\\
\(^{24}\mathrm{Mg}/^{20}\mathrm{Ne}\) & 0.1 & 1.3523 & 1746 & \(3.44\times10^{9}\)\\
\(^{12}\mathrm{C}/^{16}\mathrm{O}\) & 1 & 1.3811 & 1050 & \(2.02\times10^{10}\)\\
\(^{20}\mathrm{Ne}/^{16}\mathrm{O}\) & 1 & 1.3670 & 1431 & \(6.67\times10^{9}\)\\
\(^{24}\mathrm{Mg}/^{20}\mathrm{Ne}\) & 1 & 1.3489 & 1706 & \(3.44\times10^{9}\)\\
\end{tabular}
\end{ruledtabular}
\end{table*}

\begin{figure}
    \centering
    \includegraphics[width=\columnwidth]{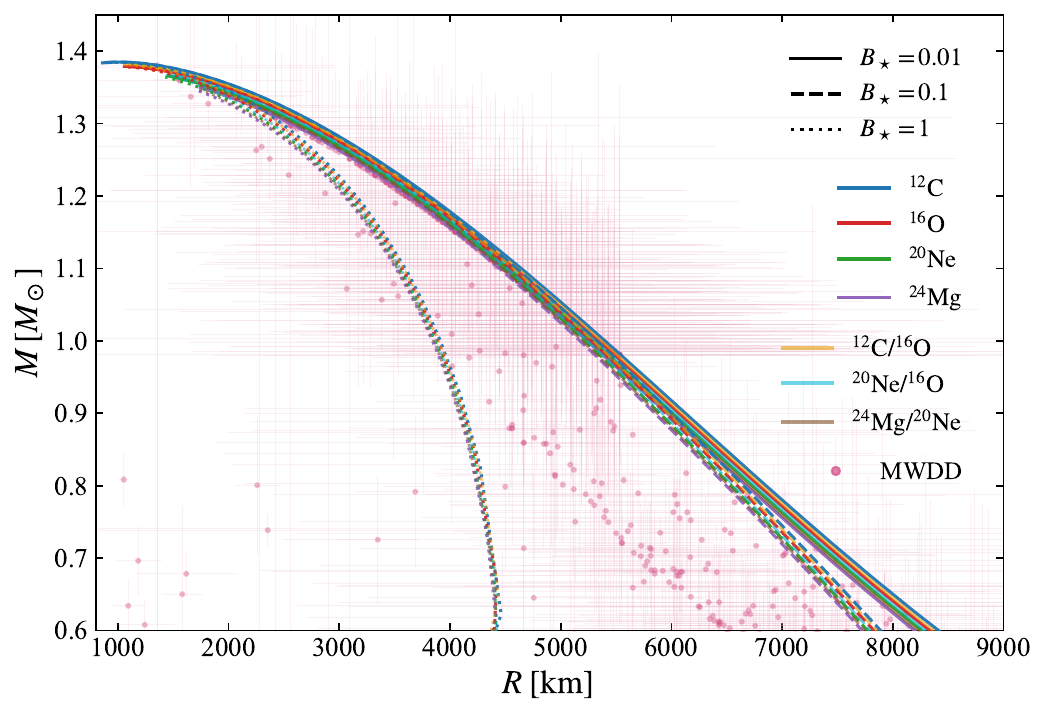}
    \caption{Exploratory observational context for the compact part of the mass versus radius plane. The points are selected from the online Montreal White Dwarf Database~\cite{MWDDOnline}, whose database methodology is described in Ref.~\cite{Dufour2016MWDD}, and are shown with the pure and mixed branches from Figs.~\ref{fig:results_pure_mr} and \ref{fig:results_mixed_mr}. The selection keeps objects in the displayed compact domain whose masses lie below the largest plotted theoretical mass at the same radius.}
    \label{fig:results_mwdd_context}
\end{figure}

As a final observational context, Fig.~\ref{fig:results_mwdd_context} compares the theoretical branches with a selected subset of WDs from the Montreal White Dwarf Database. The purpose of this comparison is limited: from the observed catalog we applied a filter to retain compact objects in the displayed mass and radius domain, while keeping the reported observational uncertainties in the plotted points. The stellar masses were taken from the catalog, and the radii were reconstructed from the measured surface gravity through \(R=\sqrt{GM/g}\), with \(g=10^{\log g}\). The radius uncertainties shown in the figure were obtained by propagating the catalog uncertainties in mass and \(\log g\). With future measurements of higher precision, this type of comparison may help assess whether magnetic pressure anisotropy can act as one of the physical mechanisms contributing to the compact radii inferred for some massive WDs.

The propagated error bars represent the catalog uncertainties in mass and $\log g$ under the reconstruction above; they do not constitute a complete treatment of atmospheric, composition, covariance, or magnetic-field systematics. Because the displayed subset is selected using the theoretical upper envelope, the comparison is theory-informed and cannot identify a strange-dwarf candidate or provide an independent statistical validation of the model.

\section{Discussion and Conclusion}\label{sec:discussion}

The turning-point behavior is used here only to screen the one-parameter equilibrium sequences. Since no radial-mode calculation is performed, the present results do not establish dynamical stability. The mixed branches also terminate at the first inverse-beta threshold and therefore do not represent complete weak-equilibrium crust sequences.

We examined whether pressure anisotropy from Landau quantization can affect equilibrium sequences of ultra-dense WDs when weak interaction thresholds and nuclear composition are treated explicitly. The calculation was formulated as an effective spherical stellar structure problem in which the radial pressure is identified with \(P_\parallel\) and the tangential stress with \(P_\perp\). Within this framework, the central question is whether the local anisotropy of the magnetized EoS can leave a visible imprint on the macroscopic relation between mass, radius, and central density.

The EoS analysis shows that the magnetic field separates the parallel and perpendicular pressure components in density intervals where the occupation of Landau levels is sparse. In those regimes the microscopic ordering \(P_\perp<P_\parallel\) becomes relevant for the anisotropic force term in the stellar structure equation. This effect is strongest as a local thermodynamic feature, while the total pressure scale remains controlled by the degenerate electron gas, the Coulomb lattice contribution, and the composition dependent weak interaction thresholds. The anisotropic pressure is therefore not an isolated correction to the EoS, but a channel through which the magnetic field can modify the radial balance of compact WD configurations.

The sequences that follow electron captures clarify why the mass versus radius plane alone is not sufficient to diagnose equilibrium stability. Although the capture following curves can appear continuous in \(M(R)\), the corresponding \(M(\rho_c)\) relation reveals extended portions with \(dM/d\rho_c<0\) after the maximum mass is reached. This behavior indicates that the post-capture branch should not be interpreted as an ordinary smooth family of stable WDs. The central density diagnostic is therefore essential for separating visually continuous branches from physically admissible equilibrium configurations in the effective model. For this reason, the pure and mixed pre-capture branches provide a more controlled comparison. The pure branches describe metastable configurations at fixed nuclear species and are terminated at the first relevant electron capture threshold. The binary \(50/50\) mixtures extend this construction to fixed composition mixtures, with the endpoint set by the first unstable component rather than by an arbitrary interpolation of the pure curves. This prescription highlights the role of composition: carbon-rich branches can extend to higher central densities, while oxygen, neon, and magnesium containing branches are limited earlier by weak interaction instabilities. The mixed sequences therefore connect the pure limits while preserving a physically motivated instability criterion.

At the macroscopic level, the most visible effect of the adopted anisotropic prescription is a shift of the stellar branches toward smaller radii. The maximum masses change only modestly across the magnetic field values considered here, but the radii, especially along the massive portions of the sequences, respond more clearly to the anisotropic pressure balance. In the effective TOV equation this behavior follows from the sign of \(2(P_t-P_r)/r\): when \(P_t=P_\perp\) and \(P_r=P_\parallel\), the inequality \(P_\perp<P_\parallel\) steepens the radial pressure gradient. The resulting compactification is therefore a consequence of the present anisotropic structure model and suggests that radius measurements may be more sensitive than maximum masses to this class of magnetic effects.

The comparison with the selected Montreal White Dwarf Database sample provides an observational context for this interpretation. The data points were not used to fit the model, and the selected compact objects should not be read as a statistical confirmation of the theoretical branches. Rather, Fig.~\ref{fig:results_mwdd_context} places the calculated sequences in the same mass and radius domain as published white dwarf estimates whose radii were reconstructed from catalog masses and surface gravities. With future measurements of higher precision, especially for massive magnetic WDs with independent constraints on \(M\), \(R\), composition, and field strength, this type of comparison may help assess whether magnetic pressure anisotropy contributes to the compact radii inferred for some objects.

Several extensions are required before this mechanism can be assessed in a fully relativistic magnetized stellar model. The present calculation uses an effective spherical description and does not solve the axisymmetric Einstein-Maxwell problem with a self-consistent magnetic field geometry. It also does not include the global Maxwell stress contribution to the spacetime, nor does it evolve the composition through a reaction network after the first weak instability is reached. Future work should therefore combine the microscopic anisotropic EoS with axisymmetric stellar structure, realistic magnetic field profiles, composition evolution, and observational samples with improved masses, radii, and magnetic diagnostics. In that setting, magnetic pressure anisotropy may become a testable contributor to the compactness of massive WDs.

\begin{acknowledgments}
E.O.~acknowledges support from the Funda\c{c}\~ao Cearense de Apoio ao Desenvolvimento Cient\'ifico e Tecnol\'ogico (FUNCAP) through grant BP6-0241-00335.01.00/25. JMZP acknowledges the financial support provided by FAPERJ under Process No.~SEI-260003/000308/2024.
\end{acknowledgments}

\end{document}